\documentclass{optica-article}

\journal{opticajournal} 

\articletype{Research Article}

\usepackage{lineno}
\usepackage{float}
\usepackage{comment}
\usepackage{graphicx}
\usepackage{subcaption}
\usepackage{booktabs}
\usepackage{enumitem}
\usepackage{setspace}       
\usepackage{rotating}
\usepackage[table]{xcolor}
\usepackage[ruled,linesnumbered]{algorithm2e}

\usepackage{xpatch}
\usepackage{tikz-cd}

\newcommand{\mypar}[1]{{\bf #1.}}

\begin{document}

\title{Inverse Reconstruction of Double Imaging Schlieren}

\author{Shubham Saxena,\authormark{1} Jo\~ao F. C. Mota\authormark{1,*}}

\address{\authormark{1}
School of Engineering and Physical Sciences, Heriot-Watt University, Edinburgh EH14 4AS, UK\\
}

\email{\authormark{*}j.mota@hw.ac.uk} 


\begin{abstract*}
Double imaging is a common artefact in off-axis single-mirror Schlieren systems. Depending on the distance between the object of interest and the mirror, together with the off-axis imaging geometry, such systems often capture images containing a superposition of the target image and a displaced ghost image, resulting in the so-called double imaging phenomenon. This not only degrades image quality, but also complicates its interpretation. To address this problem, we present a new model for ghost image displacement that takes into consideration finite-aperture support, defocus, and magnification. We then use this model to reconstruct the target image via total variation regularisation with a matrix-free alternating direction method of multipliers (ADMM) algorithm. Extensive experiments using both synthetic and real Schlieren images validate the proposed model, show the effectiveness and efficiency of our algorithm, and evaluate its performance under parameter mismatch. The proposed framework provides a computational alternative to hardware-based suppression of double imaging, enabling physics-consistent reconstruction while preserving the simplicity and compactness of off-axis single-mirror Schlieren systems.
\end{abstract*}

\section{Introduction}
Schlieren imaging has long been used to visualise refractive-index gradients in
transparent media and remains an important experimental technique in optical
flow diagnostics~\cite{settles2001schlieren, settles2017review}. 
By converting refractive-index
gradients induced by variations in density, temperature, or concentration into
measurable intensity contrast, Schlieren imaging enables visualising
otherwise invisible flow structures~\cite{settles2001schlieren}. 
Due to its
high sensitivity and non-intrusive nature, it has been used to study
compressible flows~\cite{nagata2025schlieren}, thermal
transport~\cite{luo2024thermal,tanda2014heat},
combustion~\cite{sim2021detailed,skeen2015simultaneous}, supersonic and
shock-wave flows, and high-speed
imaging~\cite{leidy2020high,faheem2021estimation}. Computational imaging
methods have further expanded the capabilities of Schlieren imaging by enabling
quantitative measurements~\cite{wildeman2018real,settles2017review, settles2022schlieren}. 

Schlieren systems typically use mirrors due to their ability to 
provide large apertures and long focal lengths while avoiding chromatic aberrations
associated with refractive optics~\cite{settles2001schlieren}.
Single-mirror off-axis configurations, in particular, use a single concave
mirror for both collimation and focusing. This not only simplifies
alignment~\cite{saxena2026water}, but also requires a small number of
components, making the entire system cost-effective and of simple
implementation.

\begin{figure}[t]
    \centering
    \includegraphics[width=0.70\textwidth]{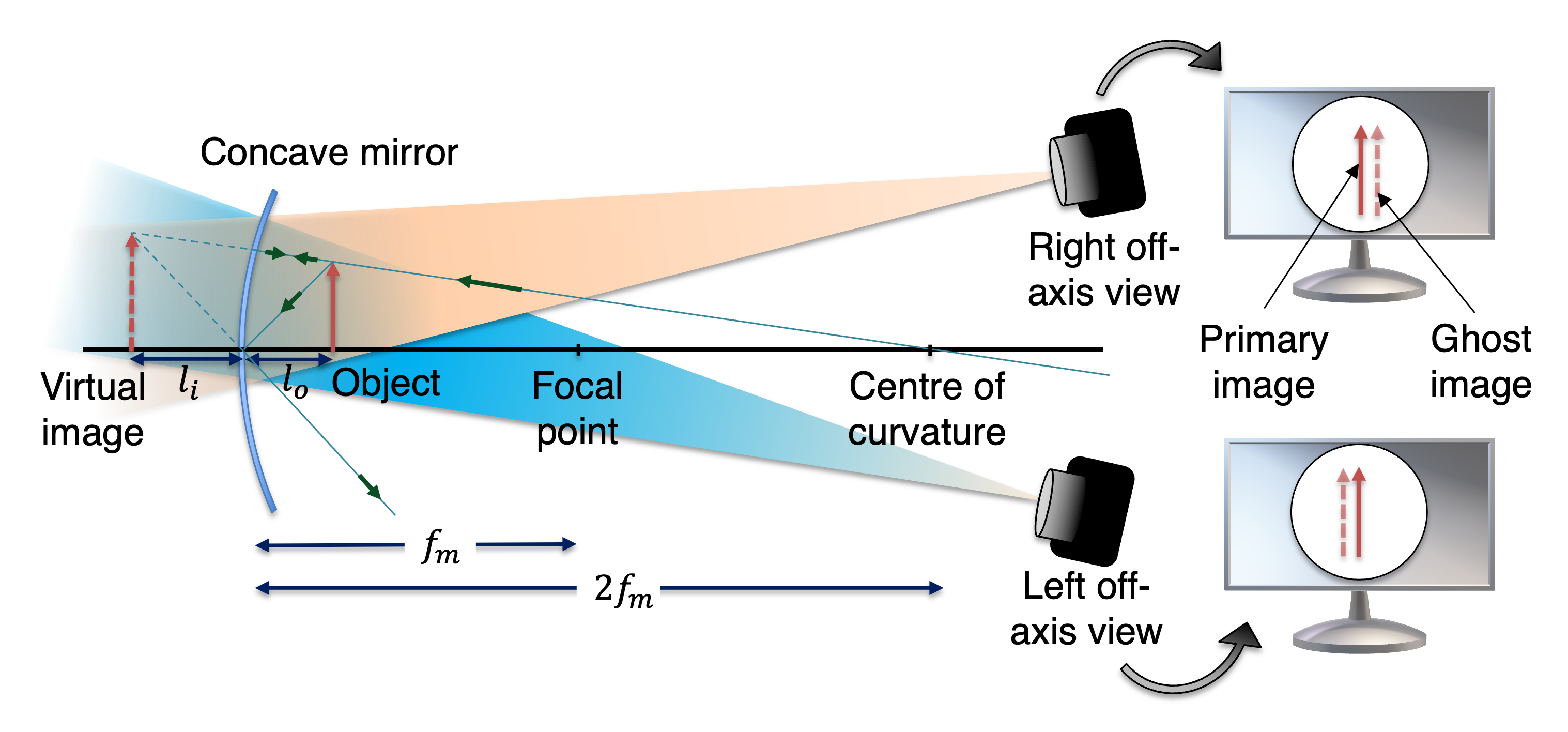}
    \caption{Double image formation in an off-axis single-mirror Schlieren system. 
        While only one camera is used in practice, we show two camera
        positions to 
        illustrate how the direction of the off-axis observation determines the
        displacement of the ghost image.
    }
    \label{fig:double_image_geometry}
\end{figure}

\mypar{The double imaging phenomenon}
Systems that use an off-axis single-mirror, however, suffer from a
recurrent artefact: \textit{double imaging}. This manifests as a faint, spatially displaced replica of the primary Schlieren image~\cite{chaloupka2014color}, similarly
to what has been observed in optical ghosting, visual reverberation, and
television deghosting~\cite{shih2015reflection,diamant2008overcoming,winters1991ghost}.
Fig.~\ref{fig:double_image_geometry} illustrates 
this phenomenon in a Schlieren setup: a concave mirror forms an image of the
light source at the cutoff plane [defined in
Fig.~\ref{fig:experimental_setup}a] and also a virtual image of the Schlieren
object behind the mirror. Refractive-index gradients in the test section
deflect the light rays, while the knife edge converts these angular deflections
into measurable intensity variations by partially blocking the light-source
image at the cutoff plane. The camera thus receives contributions from both the real Schlieren object and its virtual image, which 
appears as a displaced copy of the recorded
field~\cite{taylor1933improvements}. This displacement arises because the
illumination and viewing paths do not coincide along the Schlieren object.
The larger the separation between these paths,
the larger the displacement between the virtual and primary images~\cite{stein2022low}. 
And since the virtual image originates from a different effective object plane, it
tends to be slightly out of focus and magnified compared to the primary
image~\cite{taylor1933improvements,chaloupka2014color,davidhazy2006introduction,stein2022low,
tawil1990field}.

\mypar{Existing approaches}
A common approach to suppress double imaging is to use Schlieren configurations with on-axis coincident paths, in which the shifted secondary image overlaps with the primary image, thereby suppressing the double imaging artefact~\cite{settles2001schlieren, taylor1933improvements}.
However, such systems generally require additional optical components and complex optical alignment. Since both the illumination and viewing rays traverse the refractive-index field twice along nearly identical paths, the effective angular sensitivity increases and may become over-ranged for strong refractive-index gradients \cite{settles2001schlieren, miller2016smartphone}.
Coincident-path configurations have been implemented using beam splitters \cite{lyu2025temporally,settles2017review,biele2003point} or rod-mirror arrangements placed near the camera axis \cite{hargather2009retroreflective}.

Alternatively, single-pass setups avoid double refraction entirely, eliminating the duplicate image. Standard Z-type configurations achieve this using two parabolic mirrors aligned to minimise off-axis aberrations \cite{settles2001schlieren}, though at the expense of high component costs, a large system size, and complex alignment. 
To overcome these limitations, recent approaches use a single-mirror, single-pass V-shaped optical path, avoiding double refraction and the associated duplicate image \cite{nawrath2024optimized}. However, because this simpler configuration requires a significantly larger off-axis angle to separate the optical paths, it expands the system size and  restricts the usable field of view. The inclined illumination geometry introduces severe off-axis aberrations, particularly coma and astigmatism, which distort the horizontal and vertical foci into ellipsoids \cite{nawrath2024optimized, settles2001schlieren}. Such focal distortion prevents a uniform knife-edge cutoff, introducing directional measurement bias and uneven background illumination across the Schlieren image \cite{settles2001schlieren}.
More recently, Jian et al. \cite{jian2026comprehensive} viewed this problem as a blind deconvolution problem and used Pix2Pix~\cite{isola2017image} to suppress Schlieren artefacts, including ghosting, using images from a separate Z-type system as artefact-free references for training the network.

\mypar{Problem statement}
While double imaging can be mitigated with optical modifications, these introduce complexity, add footprint, change the field of view, and affect optical quality. Conversely, maintaining a compact, single-mirror off-axis geometry yields persistent ghost-image artefacts. We seek a computational solution that inherits the benefits of both approaches. Specifically, we adopt a simple single-mirror off-axis optical solution, and remove ghost-image artefacts by solving an inverse problem. \textit{Our goal is thus to accurately model the single-mirror off-axis setup (forward model) by capturing physical effects like aperture clipping, magnification, and ghost-image defocus, and develop associated algorithms that suppress the ghost image.}

\mypar{Proposed framework}
To the best of our knowledge, a formal physics-based mathematical framework for double imaging formation in Schlieren systems has not yet been established, with existing approaches relying primarily on optical modifications or generic image processing methods to suppress the artefact.
We introduce a physics-aware forward model for double imaging Schlieren systems. Rather than representing the ghost image as a perfectly shifted replica of the primary image, we account for 
finite-aperture effects via mirror masking operators, ghost-image defocus, and relative magnification.
Based on this model and assuming the primary image has small total variation
(TV), we then formulate an inverse problem, which we solve via the alternating direction method of multipliers (ADMM). For computational efficiency, the forward and adjoint operations are implemented in matrix-free operator form, bypassing explicit construction of large-scale matrices.

\mypar{Contributions}
We summarise our main contributions as follows:

\begin{itemize}[nosep]
\item We establish a physics-aware mathematical model for double image formation in single-mirror off-axis Schlieren systems, accounting for finite-aperture masking, ghost-image defocus, and magnification mismatch between the primary and ghost images.

\item We formulate double image suppression as a TV-regularised inverse problem and solve it efficiently using ADMM with matrix-free forward and adjoint operators.

\item We validate the proposed framework on both synthetic and experimental Schlieren measurements, demonstrating effective suppression of double imaging artefacts while preserving physically meaningful Schlieren structures.
\end{itemize}
The proposed framework provides a computational alternative to existing optical approaches 
while preserving the compact off-axis Schlieren geometry.

\mypar{Organisation}
Section~\ref{Imaging_Model} describes the proposed forward model, and 
Section~\ref{reconstruction_method} focusses on reconstruction methodology: a shift-estimation procedure and a TV-regularised ADMM reconstruction algorithm. The experimental setup is given in Section~\ref{Experimental_setup}, and experiments using synthetic and experimental data are described in Section~\ref{Results_and_discussion}. Section~\ref{conclusion} draws conclusions.

\section{Imaging model}
\label{Imaging_Model}

\subsection{Defocus- and magnification-aware forward model}
\begin{figure}[ht]
    \centering
    \includegraphics[
    width=0.94\textwidth,
    clip
    ]{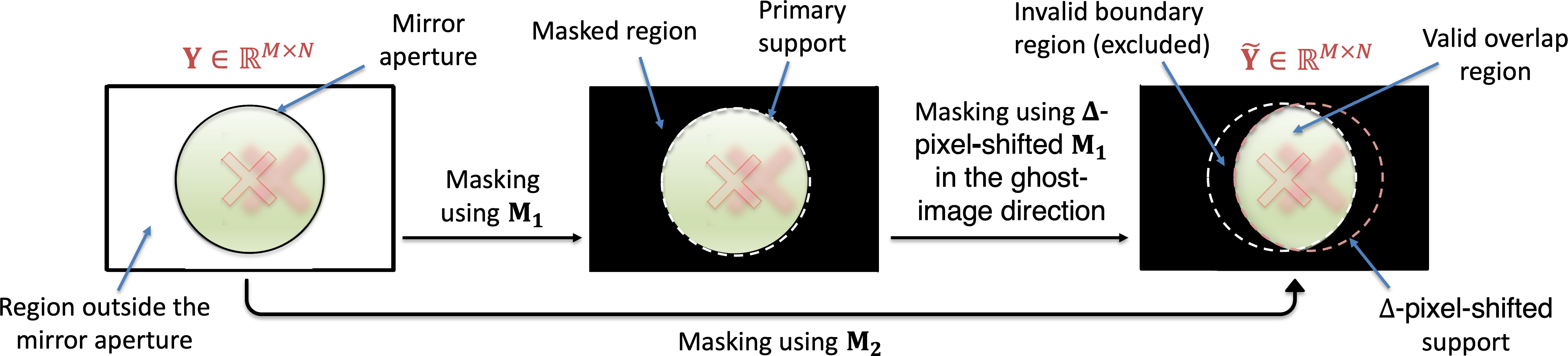}
    \caption{Illustration of the effective measurement \(\tilde{\mathbf{y}}=\mathbf{vec}(\tilde{\mathbf{Y}})\). \(\mathbf{M_1}\) restricts the recorded image \(\mathbf{Y}\) to the valid mirror aperture, while \(\mathbf{M_2}\) retains the common valid support between the primary and \(\Delta\)-pixel-shifted supports, excluding invalid boundary regions.
    }
    \label{fig:effective_measurement}
\end{figure}

\begin{figure}
    \centering
    \includegraphics[
    width=\textwidth,
    clip
    ]{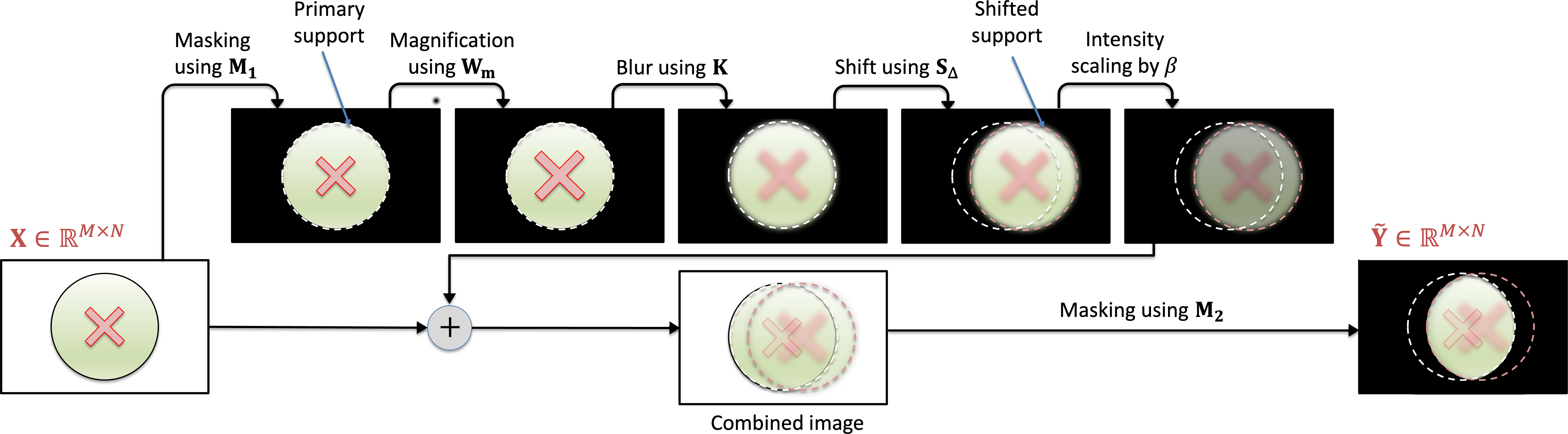}
    \caption{Illustration of the forward model and the sequence of operations used to form the primary and ghost-image contributions. The two contributions are combined and restricted to the valid overlap region by \(\mathbf{M_2}\) to obtain the effective measurement $\tilde{\mathbf{y}}$.
    }
    \label{fig:forward_model}
\end{figure}

To describe our forward model, we define
$\mathbf{X},\mathbf{Y}\in\mathbb{R}^{M\times N}$ as the primary image and the measurement, respectively, and $\mathbf{x}, \mathbf{y} \in \mathbb{R}^{MN}$ as their column-major vectorisations. As illustrated in Fig.~\ref{fig:effective_measurement}, the recorded image \(\mathbf{Y}\) contains pixels within and outside the finite mirror aperture. 
Since Schlieren measurements lie only within the mirror aperture region, we use
a mask \(\mathbf{M_1}\) to define this region. 
The ghost image is shifted by \(\Delta\) pixels relative to the primary image, so the two supports do not completely overlap. We therefore define \(\mathbf M_2\) to represent this common valid support, which excludes the non-overlapping boundary region. The support and shift operators will be described later in detail. We define the \textit{effective measurement} as 
\begin{equation}
\begin{aligned}
\tilde{\mathbf{y}}
&= \tilde{\mathbf{A}}_{M,K,W}\mathbf{x}\,,
\end{aligned}
\label{eq:forward_model}
\end{equation}
where the forward operator $\tilde{\mathbf A}_{M,K,W}(\beta) \in \mathbb{R}^{MN \times MN}$ is
\begin{equation}
\tilde{\mathbf A}_{M,K,W}(\beta)
=
\mathbf M_2
\left(
\mathbf I
+
\beta
\mathbf S_\Delta
\mathbf K
\mathbf W_m
\mathbf M_1
\right),
\label{eq:effective_operator_definition}
\end{equation}
and
$\beta\in\mathbb{R}_{+}$ denotes the relative intensity scaling of the ghost image with respect to the primary image. 
Fig.~\ref{fig:forward_model} shows the sequence of operations
in~\eqref{eq:effective_operator_definition}: the ghost contribution is formed
by applying the support masking operator \(\mathbf M_1\)
(Sec.~\ref{sec:support_shift}), followed by the magnification operator
\(\mathbf W_m\) (Sec.~\ref{sec:magnification_model}), the blur operator
\(\mathbf K\) (Sec.~\ref{sec:defocus_model}), and the shift operator \(\mathbf
S_\Delta\) (Sec.~\ref{sec:support_shift}). This produces a \(\Delta\)-pixel
displacement in the double imaging direction. The result is scaled by \(\beta\)
and added to the primary contribution, after which we apply the overlap masking
operator \(\mathbf M_2\) (Sec.~\ref{sec:support_shift}) to retain the common
valid region, yielding the effective measurement. Note the relation between the measurement and effective measurement,
$\tilde{\mathbf{y}} = \mathbf{M_2} \mathbf{y}$, visualised in Fig.~\ref{fig:effective_measurement}.
Next we define and elaborate on the above operators, each modelling a specific optical effect.
\subsubsection{Support and shift operators}
\label{sec:support_shift}

The diagonal matrix \(\mathbf{M}_1 \in \{0,1\}^{MN\times MN}\) defines a mask, with its nonzero entries indicating the spatially valid imaging region. 
This mask corresponds to the finite mirror aperture, excluding occluded or non-informative regions outside the flow field.

Spatial shifting is modelled by \(\mathbf{S}_{\Delta} \in \{0,1\}^{MN\times MN}\), which produces a displacement of \(\Delta\) pixels in the double imaging direction. We assume a non-circular boundary, with pixels shifted beyond the image boundary being discarded and the resulting incoming empty pixels set to zero.

The mask $\mathbf{M}_2 \in \{0,1\}^{MN\times MN}$, defined as $\mathbf{M}_2:=\operatorname{diag}\!\left(\mathbf{m}_1\odot\mathbf{S}_{\Delta}\mathbf{m}_1\right)$, where $\mathbf{m}_1 := \operatorname{diag}(\textbf{M}_1)$ is the diagonal of $\textbf{M}_1$ and $\odot$ the Hadamard product, excludes invalid boundary regions and restricts the observation to the common valid support shared by both image contributions.

\begin{figure}[t]
    \centering
    \includegraphics[
    width=0.84\textwidth,
    clip
    ]{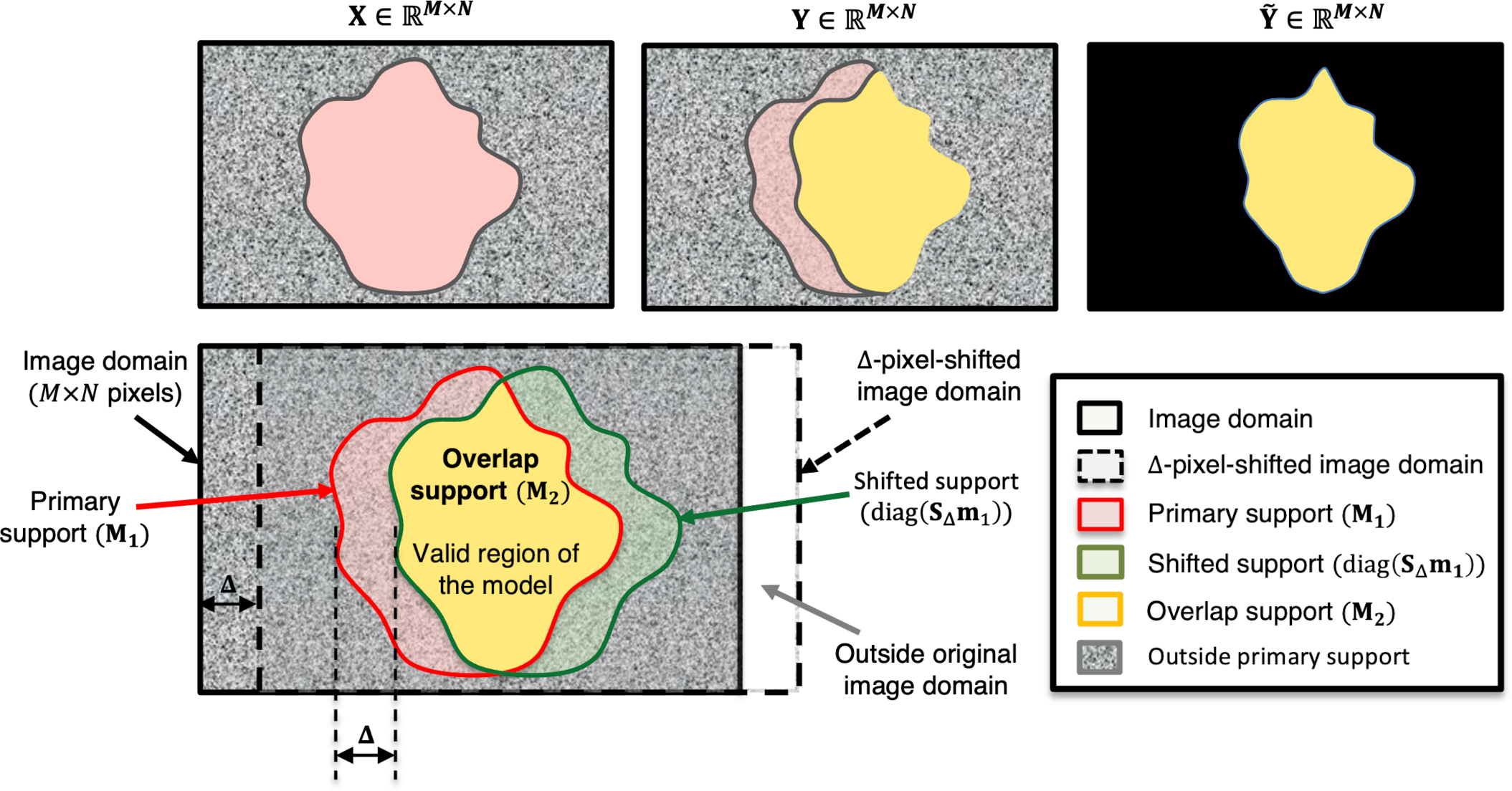}
    \caption{Illustration of the masking formulation and boundary handling for a general support geometry. The top row shows the primary image \(\mathbf X\), the measurement \(\mathbf Y\), and the effective measurement \(\tilde{\mathbf Y}\). The bottom row shows the primary support defined by \(\mathbf M_1\), its \(\Delta\)-pixel-shifted support, and their common valid region defined by \(\mathbf M_2\). 
    }
    \label{fig:masks}
\end{figure}

\mypar{Handling boundaries}
In shifted-replica inverse problems, it is important to properly handle 
boundaries to avoid stripe artefacts. 
Diamant and Schechner \cite{diamant2008overcoming} report such artefacts when inverting visual reverberations using zero-padding, and mitigate them by using mirror padding.
In our Schlieren system, 
\(\mathbf{M}_1\) and \(\mathbf{M}_2\) restrict both the forward model and the measurements to the physically valid overlapping support rather than extrapolating unknown boundary values. The proposed masking formulation is independent of the support geometry and can accommodate arbitrary valid imaging supports.
Fig.~\ref{fig:masks} 
illustrates this masking formulation for a general support geometry, 
showing how a $\Delta$-pixel shift produces partially overlapping primary and shifted supports, which $\mathbf{M}_2$ restricts to their common valid region.

\begin{figure}[t]
\centering

\begin{subfigure}[b]{0.50\textwidth}
    \centering
    \includegraphics[width=\linewidth]{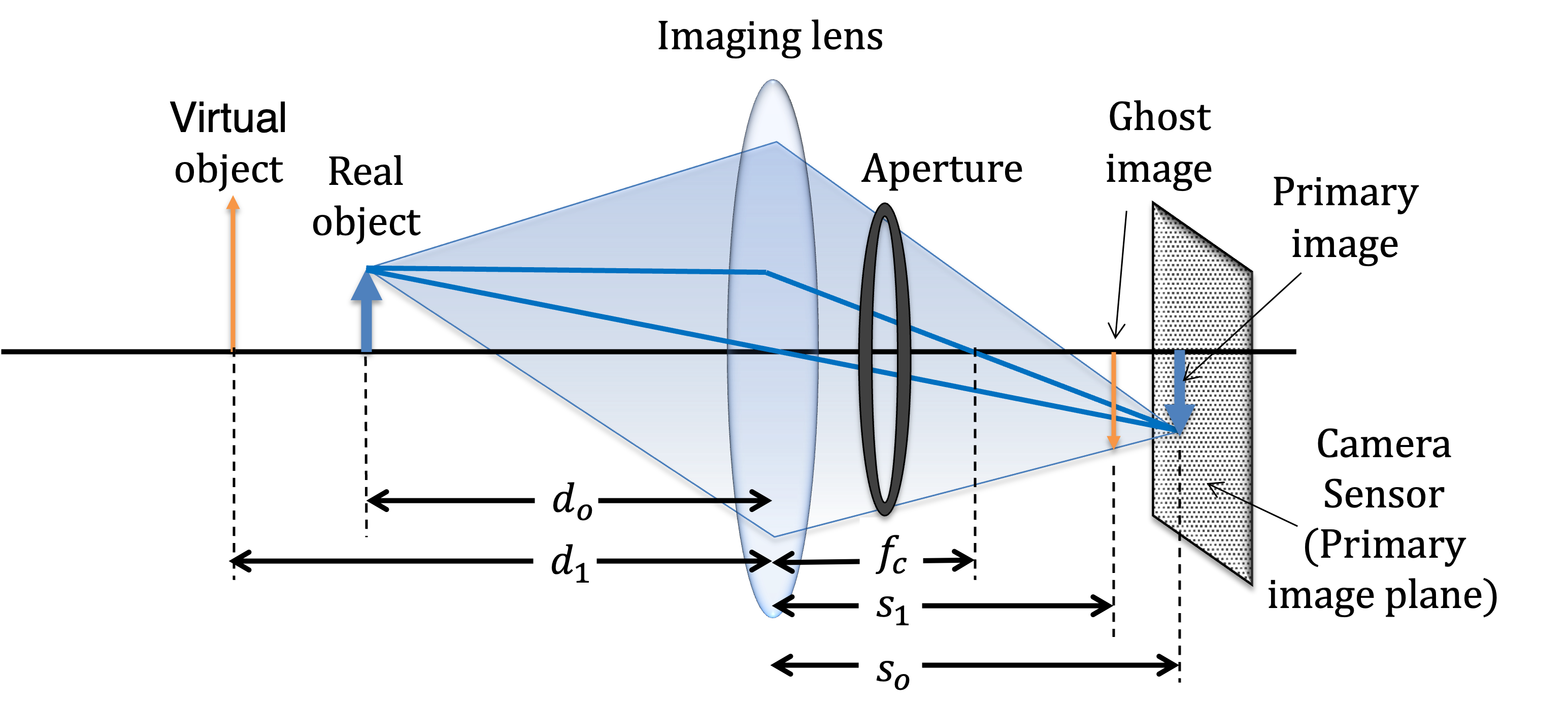}
    \caption{Primary image formation on the camera sensor from the real object plane.}
    \label{fig:ghost_geometry_a}
\end{subfigure}
\hfill
\begin{subfigure}[b]{0.49\textwidth}
    \centering
    \includegraphics[width=\linewidth]{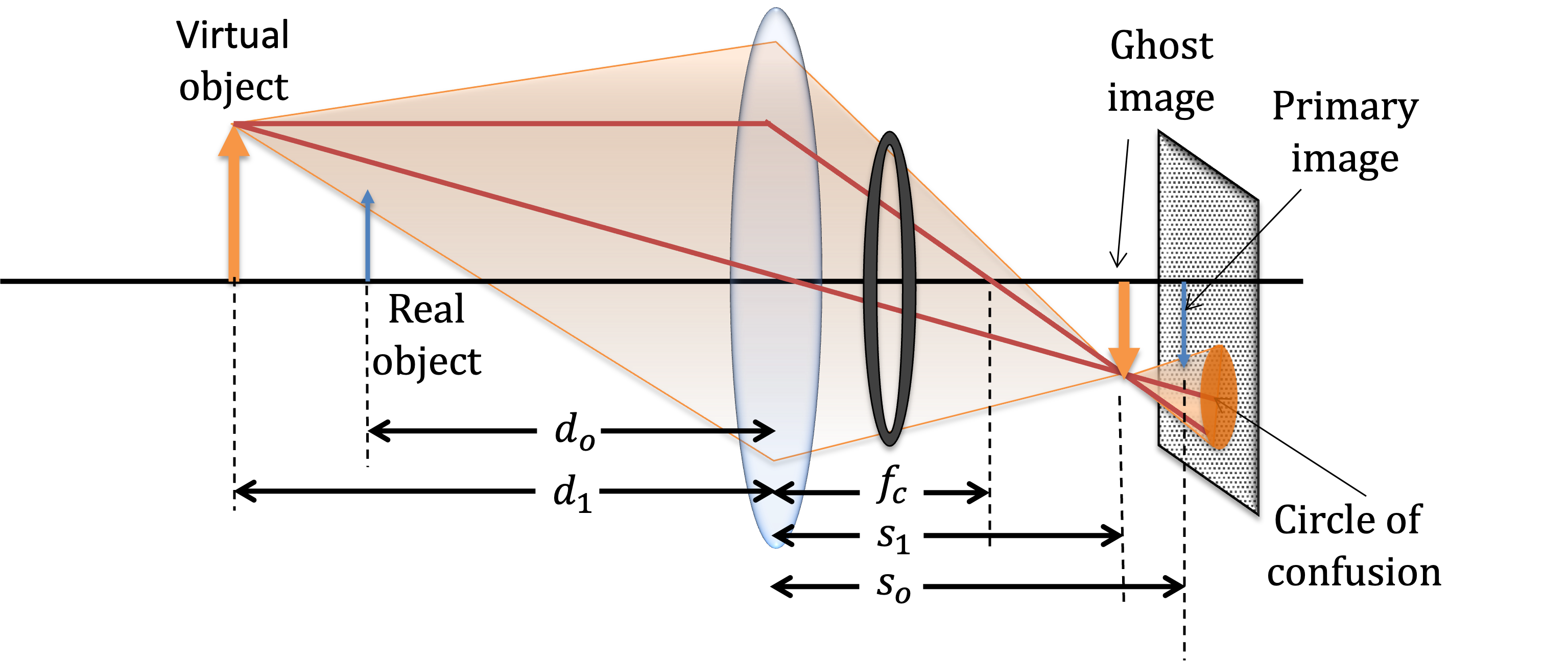}
    \caption{Secondary (ghost) image formation near the camera sensor from the virtual object plane.}
    \label{fig:ghost_geometry_b}
\end{subfigure}

\caption{
Image formation geometry in the double imaging Schlieren system. The imaging lens forms demagnified images of the real and virtual objects, while the longitudinal separation between their image planes causes the ghost image to be defocused through a circle of confusion on the sensor. (Figures are not drawn to scale.)}
\label{fig:ghost_geometry}
\end{figure}

\subsubsection{Ghost-image defocus model}
\label{sec:defocus_model}
\medskip
\noindent 
The operator
$\mathbf{K} \in \mathbb{R}^{MN \times MN}$ in~\eqref{eq:effective_operator_definition} models the blur induced by defocusing. Specifically, the 
primary and secondary image planes have different focal points, causing a focus mismatch.
Fig.~\ref{fig:ghost_geometry} shows the camera-lens image formation for the primary [Fig.~\ref{fig:ghost_geometry_a}] and ghost [Fig.~\ref{fig:ghost_geometry_b}] images.  
The corresponding object planes, located at distances \(d_0\) and \(d_1\) from the camera lens, are imaged at distinct image-plane locations \(s_0\) and \(s_1\), respectively.
Using the thin-lens equation with camera-lens focal length \(f_c\), the respective image-plane distances are

\begin{equation}
s_0 = \frac{f_c d_0}{d_0 - f_c}, \qquad s_1 = \frac{f_c d_1}{d_1 - f_c}.
\label{eq:image_planes}
\end{equation}
The ghost-image defocus is characterised by the diameter of the circle-of-confusion \cite{ray2002applied}.
The sensor is positioned at the image plane $s_0$, where the primary image is in focus. 
The ghost image is instead focused at $s_1$, and since $s_1<s_0$, its rays converge before reaching the sensor and subsequently diverge, forming a circle of confusion in the sensor plane [Fig.~\ref{fig:ghost_geometry_b}]. 
Using similar-triangle geometry, the corresponding blur-circle diameter is \(c_1=A_c\big|1-s_0/s_1\big|\) \cite{bae2007defocus}, where \(A_c:=f_c/f_{\#}\) is the camera aperture diameter and \(f_{\#}\) denotes the \(f\)-number of the camera lens. 
Using \eqref{eq:image_planes}, the blur-circle diameter becomes
\begin{equation}
c_{\mathrm{1}}=
\frac{f_c^2}{f_{\#}(d_0-f_c)}
\left|1-\frac{d_0}{d_1}\right|\,.
\label{eq:defocus_blur_diameter}
\end{equation}

When the camera lens is positioned near the centre of curvature of the mirror, \(2f_m\) [see Fig.~\ref{fig:experimental_setup}a], the distances in~\eqref{eq:defocus_blur_diameter} of the real and virtual objects to the lens are expressed as \(d_0=2f_m-l_o\) and \(d_1=2f_m+|l_i|\). Discretising \(c_1\) over a sensor with effective pixel pitch \(p_{\mathrm{eff}}=W_{\mathrm{sensor}}/N\), where \(W_{\mathrm{sensor}}\) is the sensor width and \(N\) the horizontal image dimension, gives the continuous blur radius in pixel units as \(r_1=c_1/(2p_{\mathrm{eff}})\).

\mypar{Pillbox kernel}
We model the ghost-image defocus with a normalised disk (pillbox) point spread function (PSF) kernel~\cite{banham2002digital,zheng2006extended}: \(k(\mathbf r)=1/(\pi r_1^2)\) if \(\|\mathbf r\|\leq r_1\), and \(k(\mathbf r)=0\) otherwise. The corresponding discrete convolution operator is denoted by \(\mathbf K\).
The conventional circle-of-confusion model assumes that the object of interest
emits or scatters light across the full camera aperture, which is appropriate for ordinary intensity objects, such as the candle body or a flame.
Schlieren structures, however, are not self-luminous, but arise from angular deflections of a near-collimated test beam caused by refractive-index gradients. Therefore, the conventional circle-of-confusion model overestimates the blur associated with the ghost image.

In a single-mirror Schlieren configuration, if the light source has an aperture
with diameter $d_s$ located at $2f_m$, the mirror forms an inverted image of equal diameter at the same axial distance $2f_m$ from the mirror. 
The reflected rays that appear to originate from a point in the virtual object
plane have an angular spread limited by the finite source image diameter $d_s$.
Over the longitudinal separation $l_o+|l_i|$ between the virtual and Schlieren
object planes, this angular spread produces a blur spot of diameter
$c_{s,\mathrm{obj}}=d_s(l_o+|l_i|)/(2f_m+|l_i|)$ at the Schlieren object plane
(see Appendix~\ref{app:blur_spot} for a derivation). Substituting $|l_i|=f_ml_o/(f_m-l_o)$ gives $c_{s,\mathrm{obj}}=d_sl_o/f_m$. Since the camera focuses on the Schlieren object plane, this blur spot is imaged onto the sensor with 
magnification $m_{\mathrm{p}}$, yielding a blur-spot diameter $c_2=c_{s,\mathrm{obj}}m_{\mathrm{p}}$.
Converting \(c_2\) to pixel units gives the blur radius
\begin{equation}
r_{\mathrm{2}}
=
\frac{c_{\mathrm{2}}}{2p_{\mathrm{eff}}}
=
\frac{d_s l_o\,m_p}
{2f_m p_{\mathrm{eff}}}\,,
\label{eq:source_limited_pixel_blur}
\end{equation}
which scales linearly with the source aperture diameter \(d_s\). As \(d_s \rightarrow 0\), the blur radius approaches zero, recovering the classical point-source Schlieren limit, in which the illumination is approximately collimated, and the Schlieren image remains sharply focused over an extended depth range \cite{settles2001schlieren}.
When \(2r_{\mathrm{2}}<1\) pixel, the blur footprint is smaller than 
one sensor pixel and can therefore be treated as unresolved. In this regime, the ghost-image blur operator is well approximated by the identity: \(\mathbf{K}\approx\mathbf{I}\).
Conversely, when \(2r_2 \geq 1\) pixel, for example, if one uses extended light sources~\cite{saxena2025advanced}, the blur is treated as resolved 
and is  
modelled by the pillbox 
PSF.

The former predicts blur caused by camera-lens defocus, whereas the finite source aperture limits the angular extent of the illumination. The effective blur radius is therefore \(r=\min(r_1,r_2)\), and the blur operator \(\mathbf K\) is constructed from a pillbox PSF of radius \(r\).

\subsubsection{Ghost-image magnification model}
\label{sec:magnification_model}
\medskip
\noindent 
The matrix $\mathbf W_m \in \mathbb R^{MN\times MN}$ models the relative scaling between the primary and ghost images. 
As in Fig.~\ref{fig:double_image_geometry}, the Schlieren object in a single-mirror setup is typically positioned close to the mirror, between the mirror and the focal plane. Let $f_m$ denote the mirror focal length, and $l_o$ (resp.\ $l_i$)  the distance from the mirror vertex to the object (resp.\ virtual image). The light source and camera are positioned off-axis near the centre of curvature of the mirror, approximately $2f_m$ from the mirror. For $0<l_o<f_m$, the image formed by the mirror is virtual ($l_i<0$).
Using the mirror equation \cite{pedrotti2017introduction},
$1/f_m=1/l_o+1/l_i$,
the magnification factor of the virtual image is
\begin{equation}
\label{eq:magnification}
m_{\mathrm{mirror}}
=
-\frac{l_i}{l_o}
=
\frac{f_m}{f_m-l_o}\,.
\end{equation}
The transverse camera-lens magnifications for the primary and ghost
contributions are $m_p = s_0 / d_0$ and $m_g = s_1 / d_1$, respectively.
The relative camera-lens magnification of the ghost contribution with respect to the primary field is therefore
\begin{equation} 
\label{eq:relmag}
m_{\mathrm{rel}} = \frac{m_g}{m_p} = \frac{s_1 d_0}{s_0 d_1} = \frac{d_0 - f_c}{d_1 - f_c}\,.
\end{equation}
Since the secondary object is 
a magnified version of the primary object,
the net scale factor \(m_{\rm net}\) of the ghost image on the sensor
is the product between \eqref{eq:magnification} and~\eqref{eq:relmag}:
\begin{equation}
m_{\mathrm{net}} = m_{\mathrm{mirror}} \cdot m_{\mathrm{rel}} = \left(\frac{f_m}{f_m - l_o}\right) \left(\frac{d_0 - f_c}{d_1 - f_c}\right)\,.
\label{eq:net_magnification}
\end{equation}
We incorporate this net magnification factor into the forward model in~\eqref{eq:forward_model} via \(\mathbf{W}_m \in \mathbb{R}^{MN\times MN}\), implemented as a sparse bilinear interpolation matrix that scales the image coordinates with respect to the image centre. The resulting resampling operation is linear.

\subsection{Limiting cases}
The following limiting cases demonstrate that the proposed formulation naturally reduces to simpler forward models under appropriate physical assumptions.

\mypar{Negligible defocus and magnification mismatch} 
When the longitudinal separation between the primary and secondary object planes is sufficiently small such that defocus and magnification mismatch become negligible (\(\mathbf K\to\mathbf I,\mathbf W_m\to\mathbf I\)), the proposed model in~\eqref{eq:forward_model}-\eqref{eq:effective_operator_definition} reduces to the aperture-constrained forward model
\begin{equation}
    \tilde{\mathbf y}
    =
    \tilde{\mathbf A}_M\mathbf x,
    \qquad
    \tilde{\mathbf A}_M
    =
    \mathbf M_2
    \left(
    \mathbf I
    +
    \beta \mathbf S_\Delta \mathbf M_1
    \right).
\end{equation}
In our Schlieren setup, this approximation arises when  \(l_o \ll f_m\), i.e., when the separation between the primary and secondary object planes becomes negligible.

\mypar{Baseline shift model} If we neglect aperture truncation (\(\mathbf M_1\to\mathbf I\)), the overlap mask also reduces to the identity (\(\mathbf M_2\to\mathbf I\)), meaning the effective measurement returns to the unmasked observation (\(\tilde{\mathbf{y}} \to \mathbf{y}\)). Assuming \(\mathbf K\to\mathbf I, \mathbf W_m\to\mathbf I\), the model in~\eqref{eq:forward_model}-\eqref{eq:effective_operator_definition} reduces to
\begin{equation}
\mathbf{y} = \mathbf{A}\mathbf{x}, \qquad \mathbf{A} = \mathbf{I} + \beta \mathbf{S}_{\Delta}\,.\phantom{aaaaaa}
\label{eq:operator_form_reduction}
\end{equation}
In this reduced model, we observe a
superposition of a primary image and a shifted,
intensity-scaled secondary replica. This is
equivalent to the two-pulse ghosting
model employed in image reflection and deghosting
studies, where ghost formation is described by a
shifted attenuated copy of the underlying image \cite{shih2015reflection}.
More generally, visual reverberation models describe image formation using point-spread functions composed of multiple displaced and attenuated replicas generated by successive reflections \cite{diamant2008overcoming}.

\section{Reconstruction method}
\label{reconstruction_method}

Although the matrix in~\eqref{eq:effective_operator_definition} representing our forward model is square, it is rank-deficient (e.g., $\mathbf{M}_2$ is a mask). This makes the problem of finding $\mathbf{x}$ from $\tilde{\mathbf{y}}$ ill-posed. And while several parameters of the forward model can be estimated from the experimental setup (Section~\ref{Experimental_setup}), the ghost image displacement $\Delta$ is unknown in practice. 
Ideally, \(\mathbf{x}\) and \(\Delta\) would be recovered jointly by solving
\begin{equation}
(\hat{\mathbf{x}},\hat{\Delta}) = \underset{\mathbf{x},\Delta}{\arg\min}\,\,\, \frac{1}{2} \left\| \tilde{\mathbf y} - \tilde{\mathbf A}_{M,K,W}(\Delta)\mathbf{x} \right\|_2^2 + \mathcal{R}(\mathbf{x})\,,
\label{eq:x_Delta_joint}
\end{equation}
where the first term penalises deviations from the forward model~\eqref{eq:forward_model}, and $\mathcal{R}(\mathbf{x})$ is a regularisation term encoding prior assumptions about the image. While there are several options to select $\mathcal{R}(\mathbf{x})$, from functions inducing sparsity in the wavelet or DCT domain~\cite{foucart2013mathematical} to functions whose proximal operators are neural network denoisers~\cite{venkatakrishnan2013plug, ryu2019plug}, we opted to use a total variation (TV)~\cite{osher1990feature, vella2021robust} term, $\mathcal{R}(\mathbf{x})=\lambda_{\mathrm{tv}}\|\mathbf D\mathbf{x}\|_1$, where 
$\mathbf D\in\mathbb R^{2MN\times MN}$ denotes the discrete gradient operator (vertical concatenation of horizontal and vertical differences), and $\lambda_{\mathrm{tv}}>0$ a trade-off parameter. The reason is its simplicity, lack of reliance on training data and hallucinations and, according to our experiments, better performance compared to, e.g., a DnCNN denoiser~\cite{zhang2017beyond}. Assuming that $\mathbf{x}$ has small (anisotropic) total variation (TV) encodes the fact that the primary image has a small number of pixels corresponding to edges (intensity transitions) compared to 
all the pixels.

The first term in~\eqref{eq:x_Delta_joint} makes the problem non-convex, since $\tilde{\mathbf A}_{M,K,W}$ depends on the variable \(\Delta\). However, since \(\Delta\) is discrete, we solve a set of independent problems (possibly in parallel), each with a fixed candidate shift $\Delta_{\delta}$, pre-selected using correlation analysis, as described next. 
\subsection{Shift-wise reconstruction and model selection}
\label{Shift_Estimation}

As an initial estimate, the measured image is correlated with shifted versions of itself over a range of candidate displacements. For a shift $\Delta$, we define the correlation score as 
$C(\Delta)=
\bigl(|\nabla\mathbf{y}|^\top
\mathbf{S}_{\Delta}|\nabla\mathbf{y}|\bigr)/
\bigl(\||\nabla\mathbf{y}|\|_2
\|\mathbf{S}_{\Delta}|\nabla\mathbf{y}|\|_2\bigr)$, where $|\nabla\mathbf{y}|$
is the vector containing the magnitudes of the entries of the image gradient
$\nabla\mathbf{y}$.
We compute correlations using magnitudes of image gradients rather than raw intensities
to reduce sensitivity to variations in illumination. The resulting
correlation curve provides an initial estimate of the ghost displacement and
narrows down the candidate shift range considered during reconstruction.

Let \(\{\Delta_\delta\}_{\delta=1}^{R}\) denote the set of pre-screened candidate shifts. 
For each candidate \(\Delta_\delta\), we solve~\eqref{eq:x_Delta_joint} with
fixed $\Delta = \Delta_{\delta}$ and $\lambda_{\mathrm{tv}} =
\lambda|\Omega_\delta|/(2MN)$, where $|\Omega_\delta|$ is the number of valid
pixels for candidate shift $\Delta_\delta$, and $\lambda>0$ a free parameter. 
We record the cost
$F_\delta= 1/(2|\Omega_\delta|) \left\| \tilde{\mathbf y}- \tilde{\mathbf
A}_{M,K,W}(\Delta_\delta)\hat{\mathbf x}_\delta \right\|_2^2 +
\lambda/2MN \left\| \mathbf D \hat{\mathbf x}_\delta \right\|_1$
and associated reconstruction $\hat{\mathbf{x}}_\delta$.
We then select the best estimate as
$ \hat{\delta}=\arg\min_\delta F_\delta, $
and set \(\hat{\Delta}=\Delta_{\hat{\delta}}\) and \(\hat{\mathbf{x}}=\hat{\mathbf{x}}_{\hat{\delta}}\).

\subsection{TV-regularised inverse problem}

For each 
candidate shift $\Delta_{\delta}$, we solve the convex problem that results from fixing $\Delta = \Delta_{\delta}$ in~\eqref{eq:x_Delta_joint}:
\begin{equation}
\hat{\mathbf x}_\delta = \arg\min_{\mathbf{x}}\;
\frac{1}{2}
\left\|
\tilde{\mathbf y}
-
\tilde{\mathbf A}_{M,K,W}(\Delta_\delta)\mathbf x
\right\|_2^2
+
\lambda_{\mathrm{tv}}
\|\mathbf D \mathbf x\|_1.
\label{eq:tv_reconstruction}
\end{equation}
where \(\lambda_{\rm tv}=|\Omega_\delta|\lambda/(2MN)\).  
We describe next how to solve this problem efficiently.

\begin{algorithm}
\caption{Double imaging Schlieren reconstruction framework}
\label{alg:reconstruction}

\KwIn{Measured image $\mathbf{y}$, primary support mask $\mathbf{M}_1$, blur operator $\mathbf{K}$, magnification operator $\mathbf{W}_m$, fixed ghost-image intensity coefficient $\beta$, regularisation parameter $\lambda$, ADMM parameter $\rho$.}

Estimate the candidate shifts
$\{\Delta_\delta\}_{\delta=1}^{R}$
using the correlation-based pre-screening\;

\For{$\delta=1,\ldots,R$}{

    Set $\Delta=\Delta_\delta$\;

    Construct the overlap mask $\mathbf{M}_2$ for $\Delta$\;

    Compute the effective measurement
    $\tilde{\mathbf{y}}=\mathbf{M}_2\mathbf{y}$\;

    Set $\lambda_{\mathrm{tv}}=
    \lambda|\Omega_\delta|/(2MN)$\;

    Construct the forward operator
    $\tilde{\mathbf A}_{M,K,W}$
    using $\Delta$ and $\beta$\;

    Reconstruct $\hat{\mathbf{x}}_\delta$ via ADMM
    by iterating
    \eqref{eq:x_update1}-\eqref{eq:dual_update}\;

    Evaluate the reconstruction cost $F_\delta$\;
}

Select
$\hat{\delta}=\arg\min_{\delta}F_\delta$\;

\KwOut{Optimal shift
$\hat{\Delta}=\Delta_{\hat{\delta}}$
and reconstructed image
$\hat{\mathbf{x}}=\hat{\mathbf{x}}_{\hat{\delta}}$.}

\end{algorithm}

\subsection{ADMM formulation}
We apply the alternating direction method of multipliers (ADMM)\cite{Boyd11-ADMM} to solve~\eqref{eq:tv_reconstruction}. To make it applicable to~\eqref{eq:tv_reconstruction}, we introduce a variable $\mathbf z = \mathbf D\mathbf x$, reformulating it as
\begin{align}
\min_{\mathbf{x},\mathbf z}
\quad
&
\frac{1}{2}
\left\|
\tilde{\mathbf y}
-
\tilde{\mathbf A}_{M,K,W}\mathbf x
\right\|_2^2
+
\lambda_{\mathrm{tv}}
\|\mathbf z\|_1
\notag
\\
\text{s.t.}
\quad
&
\mathbf D \mathbf x = \mathbf z.
\label{eq:tv_split_problem}
\end{align}
Associating the dual variable $\boldsymbol{\nu}$ with its constraint, the augmented Lagrangian 
of~\eqref{eq:tv_split_problem} is
\begin{equation}
\mathcal{L}_{\rho}(\mathbf{x},\mathbf{z};\boldsymbol{\nu})
=
\frac{1}{2}
\|
\tilde{\mathbf y}
-
\tilde{\mathbf A}_{M,K,W}\mathbf x
\|_2^2
+
\lambda_{\mathrm{tv}}
\|\mathbf z\|_1
+
\boldsymbol{\nu}^{\top}
(
\mathbf D\mathbf x-\mathbf z
)
+
\frac{\rho}{2}
\|
\mathbf D\mathbf x-\mathbf z
\|_2^2,
\label{eq:augmented_lagrangian}
\end{equation}
where $\rho>0$.
ADMM consists of minimising~\eqref{eq:augmented_lagrangian} with respect to $\mathbf{x}$, then with respect to $\mathbf{z}$, and then updating the dual variable with a gradient ascent step. 

\mypar{Updating $\mathbf{x}$} 
At each iteration $t$, we compute $\mathbf{x}^{(t+1)}$ by minimising~\eqref{eq:augmented_lagrangian} with respect to $\mathbf{x}$:
\begin{equation*}
\mathbf{x}^{(t+1)}
=
\arg\min_{\mathbf{x}}
\;
\frac{1}{2}
\|
\tilde{\mathbf y}
-
\tilde{\mathbf A}_{M,K,W}\mathbf x
\|_2^2
+
(\mathbf D^{\top}\boldsymbol{\nu}^{(t)})^{\top}\mathbf x
+
\frac{\rho}{2}
\|
\mathbf D\mathbf x-\mathbf z^{(t)}
\|_2^2\,,
\end{equation*}
which boils down to solving a linear system:
\begin{equation}
\resizebox{0.92\textwidth}{!}{$
\mathbf{x}^{(t+1)}
=
\left[
\tilde{\mathbf A}_{M,K,W}^{\top}
\tilde{\mathbf A}_{M,K,W}
+
\rho\,\mathbf D^{\top}\mathbf D
\right]^{-1}
\left[
\tilde{\mathbf A}_{M,K,W}^{\top}\tilde{\mathbf y}
+
\mathbf D^{\top}
\bigl(
\rho\,\mathbf z^{(t)}
-
\boldsymbol{\nu}^{(t)}
\bigr)
\right]
$},
\label{eq:x_update1}
\end{equation}
which we do using the conjugate gradient method. The adjoint operator associated with $\tilde{\mathbf A}_{M,K,W}$ is
$\tilde{\mathbf A}_{M,K,W}^{\top}\mathbf r
=
\mathbf M_2\mathbf r
+
\beta\,\mathbf M_1\,
\mathbf W_m^{\top}\mathbf K^{\top}\mathbf S_\Delta^{\top}
\mathbf M_2\mathbf r$,
where $\mathbf S_\Delta^{\top}$ denotes the adjoint shift operator and $\mathbf K^{\top}$ is the adjoint blur operator. For symmetric kernels, such as pillbox or Gaussian blur, $\mathbf K^{\top}=\mathbf K$.

\mypar{Updating $\textbf{z}$}
To update $\mathbf{z}$, we fix $\mathbf{x}=\mathbf{x}^{(t+1)}$, and minimise~\eqref{eq:augmented_lagrangian} with respect to $\mathbf{z}$:
\begin{equation*}
\mathbf z^{(t+1)}
=
\arg\min_{\mathbf z}
\;
\lambda_{\mathrm{tv}}\|\mathbf z\|_1
-
{\boldsymbol{\nu}^{(t)}}^{\top}\mathbf z
+
\frac{\rho}{2}
\|\mathbf z-\mathbf D\mathbf x^{(t+1)}\|_2^2\,,
\label{eq:z_subproblem}
\end{equation*}
whose solution can be computed in closed-form via the soft-thresholding operation:
\begin{equation}
\mathbf z^{(t+1)}
=
\mathcal S_{\lambda_{\mathrm{tv}}/\rho}
\left(
\mathbf D\mathbf x^{(t+1)}
+
\frac{\boldsymbol{\nu}^{(t)}}{\rho}
\right),
\label{eq:z_update}
\end{equation}
where $\mathcal S_\tau(v):=\operatorname{sign}(v)\max(|v|-\tau,0)$.

\mypar{Updating $\mathbf{\nu}$}
After computing $\mathbf{x}^{(t+1)}$ and $\mathbf{z}^{(t+1)}$, we update $\nu$ as
\begin{equation}
\boldsymbol{\nu}^{(t+1)}
=
\boldsymbol{\nu}^{(t)}
+
\rho
\left(
\mathbf D\mathbf x^{(t+1)} 
-
\mathbf z^{(t+1)}
\right).
\label{eq:dual_update}
\end{equation}

We iterate these updates until the primal and dual residuals,
$\mathbf{r}^{(t+1)}
=
\mathbf D\mathbf{x}^{(t+1)}-\mathbf{z}^{(t+1)}$
and
$\mathbf{s}^{(t+1)}
=
\rho \mathbf D^{T}
\left(\mathbf{z}^{(t+1)}-\mathbf{z}^{(t)}\right),
$ satisfy the convergence tolerances, with \(\rho\) adaptively updated by residual balancing; see details in~\cite{Boyd11-ADMM}.

\subsection{Reconstruction algorithm}
Algorithm~\ref{alg:reconstruction} summarises the overall reconstruction procedure. The candidate shifts are first identified using the correlation-based pre-screening described in Sec.~\ref{Shift_Estimation}. For each candidate shift, the image is reconstructed using ADMM with \(\beta\) fixed. The corresponding cost $F_\delta$ is then evaluated for each candidate, and the solution with the lowest cost is selected.

\section{Experimental setup and model}
\label{Experimental_setup}
\begin{figure}[th]
    \centering
\begin{subfigure}[b]{0.95\textwidth}
    \centering
    \includegraphics[
    width=0.98\linewidth,
    clip
    ]{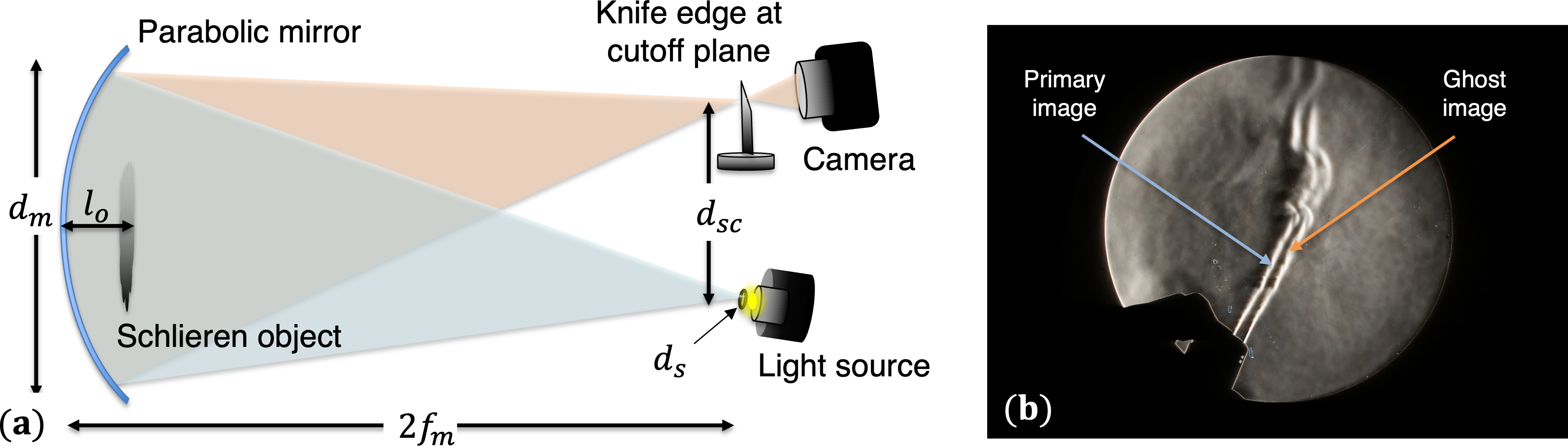}
\end{subfigure}
\caption{Experimental setup and measurement. (a) Off-axis single-mirror Schlieren system. (b) Raw measurement showing the primary and shifted ghost images.
}
\label{fig:experimental_setup}
\end{figure}

We validated the proposed reconstruction framework on 
an off-axis single-mirror Schlieren system. The experimental configuration and an example raw measurement are shown in Fig.~\ref{fig:experimental_setup}. We used a 203~mm diameter parabolic mirror with focal length $f_m=1.2$~m.
The illumination source, consisting of a pinhole aperture with diameter
$d_s=400~\mu\mathrm{m}$, 
was positioned near the centre of curvature of the mirror.
A vertical knife-edge placed at the source image plane in front of the camera lens provided the Schlieren cutoff.
We used a full-frame Sony $\alpha$7 IV camera 
with a pixel pitch of $p = 5.12\,\mu$m, 
paired with a Sony 24-105 mm lens set to \(f_c=105\,\mathrm{mm}\) and $f_{\#} = 4$,
placed along the off-axis viewing direction near the centre of curvature of the mirror.
To compute the net magnification factor in~\eqref{eq:net_magnification}, we replace $d_0 = 2f_m - l_o$ and $d_1 = 2f_m + |l_i|$, yielding
\begin{equation}
m_{\mathrm{net}}
=
\frac{(2f_m-f_c-l_o)}
{(2f_m-f_c-l_o)+\frac{f_c l_o}{f_m}}.
\label{Eq.net_relative_magnification}
\end{equation}
When \(l_o,f_c\ll f_m\), we have \(
2f_m-f_c-l_o
\gg
f_c l_o/f_m
\), and thus 
\(m_{\mathrm{net}}\approx1\). In our experiments, \(0<l_o<180~\mathrm{mm}\) which, from~\eqref{Eq.net_relative_magnification}, gives \(0.993<m_{\mathrm{net}}<1\). This justifies approximating the magnification operator by the identity: \(\mathbf{W}_m\approx\mathbf{I}\).

We compute the source-limited blur radius using~\eqref{eq:source_limited_pixel_blur}.
Although the physical sensor width is fixed  \((W_{\mathrm{sensor}}=35.9~\mathrm{mm}\)), the image resolution may vary due to downsampling.
To preserve scale invariance, the effective pixel pitch is updated as
\(p_{\mathrm{eff}}=W_{\mathrm{sensor}}/N\), where \(N\) is the horizontal image resolution.
Using the experimental parameters \(f_m=1200~\mathrm{mm}\),
\(f_{\mathrm{c}}=105~\mathrm{mm}\), and \(d_s=0.4~\mathrm{mm}\), the limiting
condition \(2r_2=1\) pixel, evaluated via~\eqref{eq:source_limited_pixel_blur},
corresponds to \(l_o\approx293~\mathrm{mm}\) for the full-resolution image
\((N=7008)\), and \(l_o\approx891~\mathrm{mm}\) for the downsampled reconstruction resolution \((N=1616)\).

All experiments were performed with
$l_o\le180$ mm,
which provides sufficient working space for introducing typical Schlieren test objects while avoiding close proximity to the mirror surface.
Consequently, the source-limited blur remains sub-pixel throughout the experiments ($\mathbf K\approx\mathbf I$).

\section{Results and discussion}
\label{Results_and_discussion}
\noindent\textbf{Synthetic validation.}
We used five standard benchmark images from the USC-SIPI Image Database (House,
Mandrill, Peppers, Boat, and Airplane) \cite{weber2018sipi}, and two images acquired by us (Cat and Swan).
These synthetic experiments enabled us to validate the proposed model~\eqref{eq:forward_model} and reconstruction Algorithm~\ref{alg:reconstruction}, under different blur and magnification factors, and study their performance when there is a mismatch between the model used to generate the measurements and the model assumed by the algorithm.
We evaluated reconstruction via the peak signal-to-noise ratio (PSNR), structural similarity index measure (SSIM)~\cite{wang2004image}, and learned perceptual image patch similarity (LPIPS)~\cite{zhang2018unreasonable}.
In all experiments, synthetic measurements were generated using a fixed ghost-image shift of $\Delta=20$ pixels and ghost-image intensity coefficient $\beta=0.8$. 
The same value of $\beta$ was used during reconstruction;
mismatch in $\beta$ was not evaluated.

\begin{table}[th]
\centering

\renewcommand{\arraystretch}{0.87}
\setlength{\tabcolsep}{7pt}

\resizebox{0.88\textwidth}{!}{%
\begin{tabular}{cccc@{\hspace{8pt}}|cccc}
\toprule

\multicolumn{2}{c}{Data model} &
\multicolumn{2}{c}{Reconst. model} &
\multicolumn{3}{c}{Reconstruction quality} & Mean Runtime
\\

\cmidrule(lr){1-2}
\cmidrule(lr){3-4}
\cmidrule(lr){5-7}

$r$ &
$m(\%)$ &
$r$ &
$m(\%)$ &
Mean PSNR &
Mean SSIM &
Mean LPIPS &
(s) \\

\midrule

0 & 0 & 0 & 0 & \textbf{36.46 $\pm$ 5.53} & 0.9713 $\pm$ 0.0141 & 0.0208 $\pm$ 0.0096 & 043.88\\

\rowcolor{black!8}
2 & 0 & 2 & 0 & 36.08 $\pm$ 5.96 & 0.9744 $\pm$ 0.0137 & 0.0242 $\pm$ 0.0061 & 074.85\\

0 & 2 & 0 & 2 & 35.70 $\pm$ 5.76 & 0.9754 $\pm$ 0.0114 & \textbf{0.0182 $\pm$ 0.0059} & 034.55\\

\rowcolor{black!8}
2 & 2 & 2 & 2 & 36.13 $\pm$ 6.02 & 0.9756 $\pm$ 0.0120 & 0.0228 $\pm$ 0.0061 & 077.55\\

0 & 5 & 0 & 5 &35.65 $\pm$ 5.87 & 0.9760 $\pm$ 0.0104 & 0.0208 $\pm$ 0.0055 & 041.73\\

\rowcolor{black!8}
6 & 0 & 6 & 0 & 35.78 $\pm$ 5.86 & 0.9766 $\pm$ 0.0109 & 0.0260 $\pm$ 0.0061 & 248.17\\

0 & 1 & 0 & 1 & 36.20 $\pm$ 5.95 & 0.9767 $\pm$ 0.0113 & 0.0194 $\pm$ 0.0061 & 036.25\\

\rowcolor{black!8}
3 & 1 & 3 & 1 & 35.94 $\pm$ 5.90 & 0.9766 $\pm$ 0.0110 & 0.0253 $\pm$ 0.0060 & 116.93\\

4 & 7 & 4 & 7 & 35.30 $\pm$ 5.88 & \textbf{0.9776 $\pm$ 0.0102} & 0.0252 $\pm$ 0.0061 & 153.90\\

\midrule

\rowcolor{black!8}
2 & 0 & 0 & 0 & 28.69 $\pm$ 4.40 & 0.8407 $\pm$ 0.0900 & 0.1470 $\pm$ 0.0572 & 060.29\\

0 & 2 & 0 & 0 & 20.48 $\pm$ 3.52 & 0.6136 $\pm$ 0.1152 & 0.3079 $\pm$ 0.0614 & 059.62\\

\rowcolor{black!8}
2 & 2 & 0 & 0 & 21.13 $\pm$ 3.56 & 0.6441 $\pm$ 0.1123 & 0.2862 $\pm$ 0.0488 & 065.23\\

2 & 2 & 2 & 0 & 23.82 $\pm$ 3.24 & 0.7877 $\pm$ 0.0692 & 0.1758 $\pm$ 0.0367 & 087.66\\

\rowcolor{black!8}
2 & 2 & 0 & 2 & 29.04 $\pm$ 2.51 & 0.9004 $\pm$ 0.0359 & 0.1023 $\pm$ 0.0207 & 045.74\\

3 & 1 & 0 & 0 & 24.15 $\pm$ 3.70 & 0.7267 $\pm$ 0.0943 & 0.2332 $\pm$ 0.0480 & 060.30\\

\rowcolor{black!8}
4 & 5 & 2 & 0 & 19.76 $\pm$ 2.70 & 0.7061 $\pm$ 0.0966 & 0.2497 $\pm$ 0.0477 & 091.25\\

3 & 6 & 0 & 3 & 19.68 $\pm$ 2.69 & 0.6441 $\pm$ 0.0899 & 0.2956 $\pm$ 0.0416 & 051.66\\

\rowcolor{black!8}
4 & 3 & 1 & 7 & 19.44 $\pm$ 2.43 & 0.6757 $\pm$ 0.0846 & 0.2810 $\pm$ 0.0488 & 070.24 \\

\bottomrule
\end{tabular}
}

\caption{Reconstruction performance (mean $\pm$ standard deviation) over seven test images for different blur radius $r$ and magnification $m(\%)$ factors in data generation and reconstruction models. The upper and lower blocks correspond, respectively, to matched and mismatched data-generation/reconstruction models. Best values in \textbf{bold}. 
}
\label{tab:synthetic}
\end{table}
Table~\ref{tab:synthetic} reports the results as the mean and population standard deviation across the seven images.
For the matched-model experiments (upper half of Table~\ref{tab:synthetic}), high reconstruction quality is consistently achieved across all benchmark images, with mean PSNR exceeding 35~dB, mean SSIM above 0.97, and mean LPIPS below 0.03. In contrast, blur and magnification mismatch between measurements and the reconstruction model (lower half of Table~\ref{tab:synthetic}) substantially degrade performance, reducing PSNR by 15+~dB, lowering the mean SSIM below 0.65, and increasing the mean LPIPS above 0.2 for several test cases. 
These results demonstrate the importance of accurately modelling blur and magnification for high-quality reconstruction.

While our setup satisfies $\mathbf{K} \approx \mathbf{I}$ and $\mathbf{W}_m \approx \mathbf{I}$, these approximations may not hold across other Schlieren configurations. Extended-source Schlieren systems ($d_s$) or setups with larger $l_o$ can produce resolvable source-limited blur ($2r_2 \ge 1$) [cf.\ \eqref{eq:source_limited_pixel_blur}], requiring a non-identity blur operator $\mathbf{K}$. Similarly, $m_{\mathrm{net}} \ll 1$ [cf.\ \eqref{Eq.net_relative_magnification}] for large \(l_o/f_m\) ratios (e.g., shorter-focal-length mirrors or larger object distances, especially when combined with large focal lengths), requiring spatial resampling via $\mathbf{W}_m$. As quantified by the model-mismatch evaluations in Table~\ref{tab:synthetic}, neglecting these effects degrades reconstruction significantly.

\begin{figure}[th]
    \centering
    \includegraphics[width=0.7\textwidth]{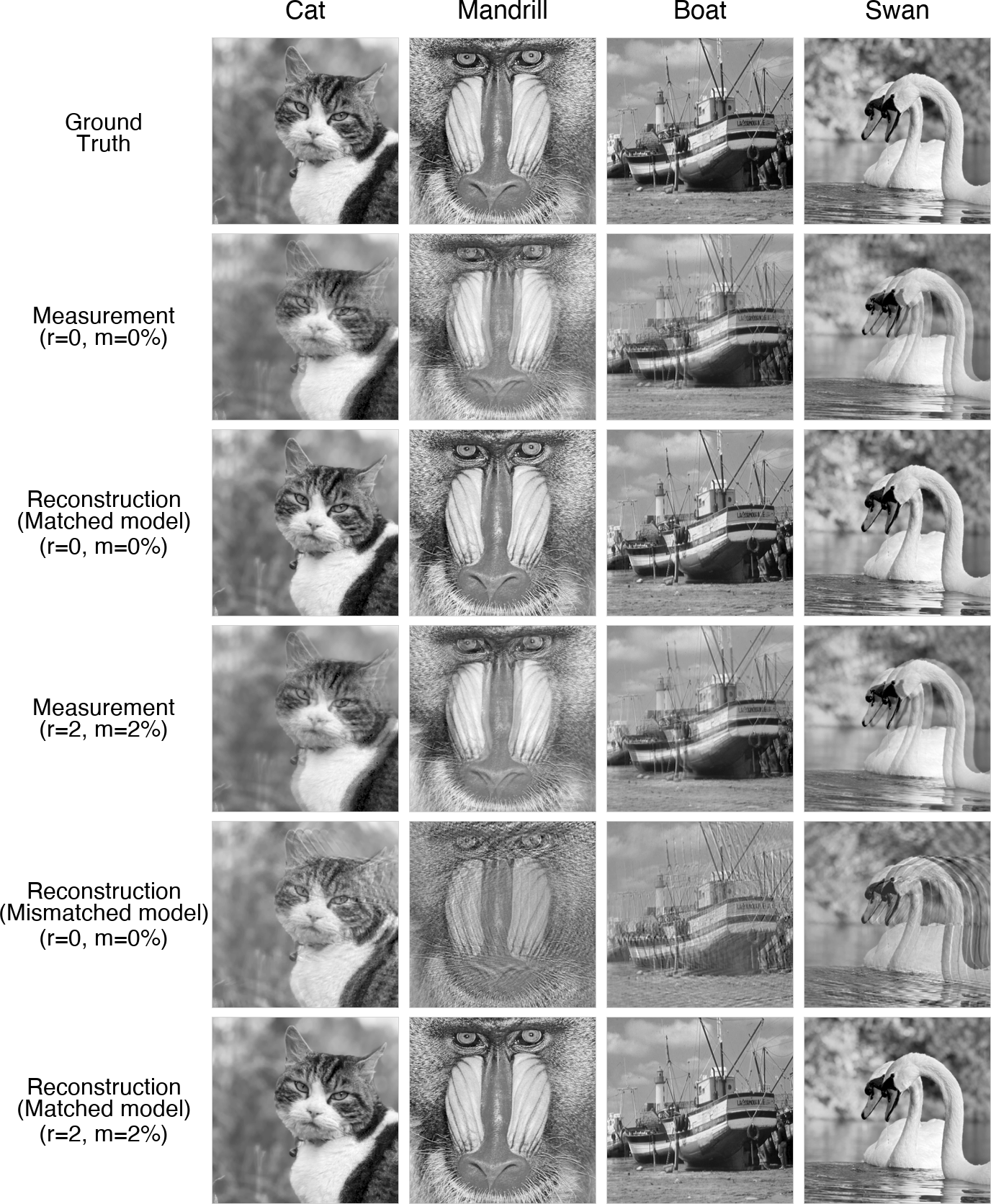}
    \caption{Reconstructions of Cat, Mandrill, Boat, and Swan test images ($512\times512$) under different combinations of blur radius ($r$) and magnification ($m\%$).  First row: ground truth. Rows~2--3: matched baseline case ($r=0$, $m=0\%$). Rows~4--5: illustration of degradation caused by model mismatch. Final row: matched case reconstruction.
    }
    \label{fig:synthetic4}
\end{figure}

These quantitative results are visually illustrated in Fig.~\ref{fig:synthetic4} for specific reconstruction examples. Matched blur and magnification parameters during reconstruction preserve fine textures and structural details. Model mismatch, however, leaves residual ghost structures and noticeable blurring, particularly around high-contrast edges.
Using the matched forward model largely suppresses these artefacts, yielding reconstructions that closely resemble the ground truth.

\mypar{Real data reconstruction}
We also evaluated our framework on real Schlieren images. Unlike the synthetic experiments, where we specified the parameters of the forward-model, here the parameters were determined from the physical characteristics of the imaging system. As discussed in Section~\ref{Experimental_setup}, the source-limited blur diameter is below one pixel. Consequently, the ghost-image blur is not expected to be resolved by the camera sensor, allowing the blur operator to be approximated by the identity. Similarly, the magnification factor between primary and ghost images is negligible for the present optical configuration.  
We therefore used a simplified forward model, with a zero blur radius ($r=0$) and no magnification mismatch ($m=0\%$).

\begin{figure}[th]
\centering
\includegraphics[
    width=0.69\textwidth,
    clip
    ]
{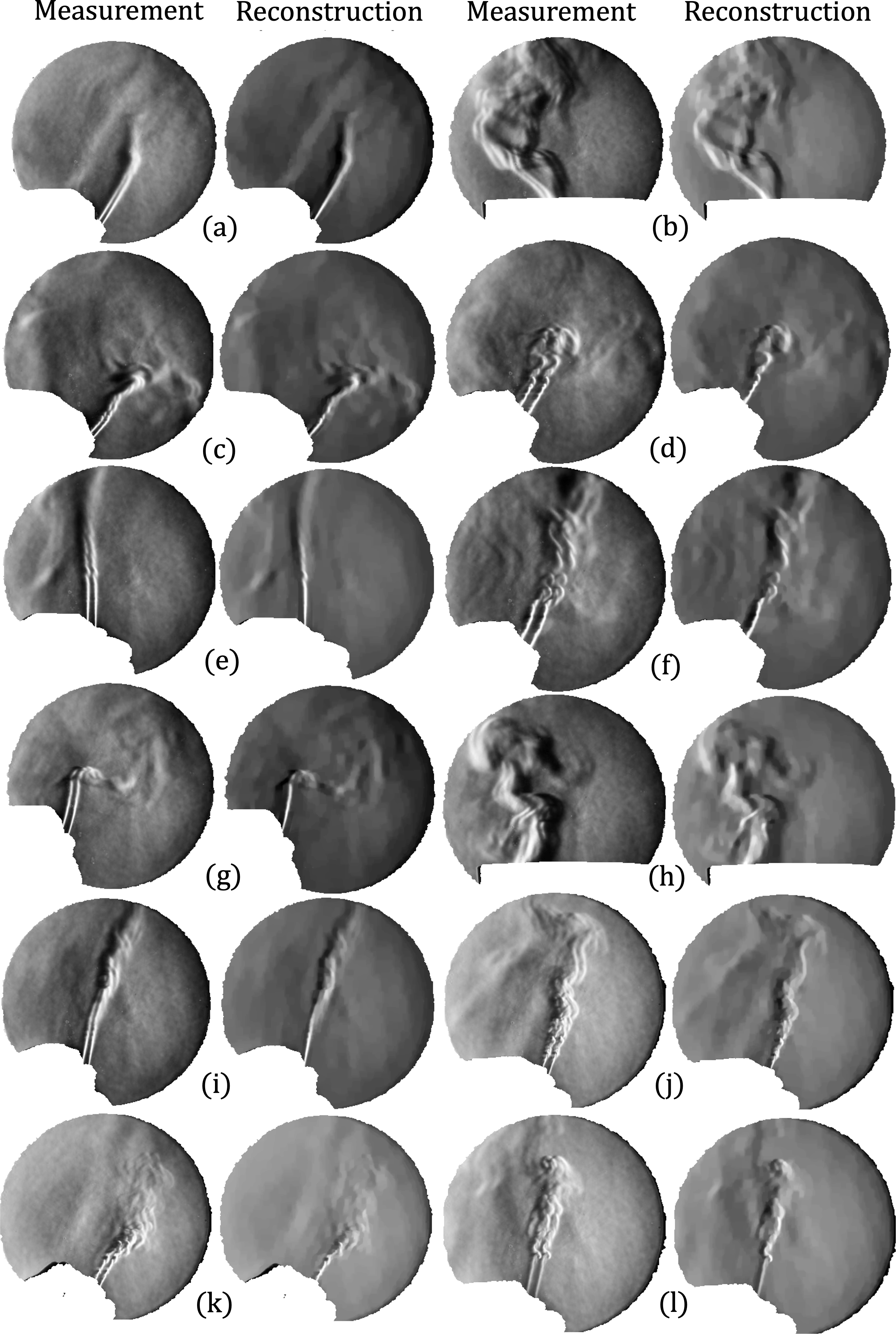}
\caption{Reconstruction results for twelve experimental Schlieren datasets, labelled (a)-(l), using \(378\times390\)-pixel regions of interest. 
Columns 1 and 3 show the measured double imaging Schlieren images containing background camera noise and dual-image ghosting artefacts. Columns 2 and 4 show the corresponding reconstructed primary images obtained using the proposed inverse reconstruction framework. Pixels outside the valid overlap support are omitted for visualisation. Across the diverse flow fields, the proposed method consistently suppresses the secondary image contribution and background noise while recovering the dominant Schlieren structures.}
\label{real_data_results}
\end{figure}
We first estimated the lateral displacement of the ghost image using the procedure of Section~\ref{Shift_Estimation}.
In our single-mirror imaging geometry, both the primary and ghost image
formation paths undergo a single reflection from the concave mirror. Hence,
neglecting any additional differential attenuation between the two optical
paths, we assumed the ghost-image intensity coefficient $\beta$ to be one. The estimated shift and fixed value of $\beta$ were then incorporated into the proposed ADMM reconstruction framework to recover the underlying primary image.

The measured images in Fig.~\ref{real_data_results} exhibit overlapping primary and secondary Schlieren structures arising from the double imaging configuration. Across all twelve experimental images, the proposed inverse framework substantially suppresses the secondary image contribution while recovering the primary Schlieren structures.
Reconstruction quality remains stable across different flow morphologies. For
the relatively laminar structures shown in Fig.~\ref{real_data_results}a
and~\ref{real_data_results}e, the duplicated boundaries are recovered as
well-defined single interfaces. For the more complex turbulent plume structures
in Fig.~\ref{real_data_results}b,~\ref{real_data_results}h,
and~\ref{real_data_results}j, our reconstruction preserves the dominant flow
features and suppresses the secondary image contribution without degrading the high-frequency features of the central convective cores.
In addition to suppressing the secondary image, the reconstruction also produces a spatially uniform background, thereby improving the visibility of the Schlieren structures.
Fine-scale turbulent structures remain visible after reconstruction, although the weakest features are slightly smoothed due to the TV regularisation [Fig.~\ref{real_data_results}k].
Overall, these results demonstrate consistent reconstruction across different experimental flow conditions without requiring explicit calibration of scene-dependent parameters. 

These results demonstrate that double imaging artefacts can be effectively addressed computationally using a physics-consistent forward model rather than through additional optical components. Although our framework estimates the ghost-image displacement, it relies on prior knowledge of the blur and magnification parameters and assumes a fixed ghost-image intensity coefficient. These assumptions are appropriate for the single-mirror Schlieren configuration considered here. When these quantities are unknown or cannot be estimated independently, reconstruction quality may be affected.

\section{Conclusions and future work}
\label{conclusion}
We proposed a physics-aware inverse reconstruction framework for double imaging Schlieren systems that models ghost-image displacement, finite-aperture support, defocus, and magnification mismatch in the forward model. Synthetic experiments demonstrate that accurate modelling of blur and magnification is essential for high-quality reconstruction, while experimental results show effective suppression of double imaging artefacts and recovery of the dominant Schlieren structures across diverse flow conditions. Unlike existing approaches that rely on optical modifications to avoid double imaging, the proposed framework provides a computational solution that preserves the simplicity and compactness of off-axis single-mirror Schlieren systems while improving image quality through physics-consistent reconstruction.
A natural extension of the present framework is the joint estimation of these forward-model parameters during reconstruction, which can be investigated in future work. More sophisticated reconstruction algorithms could also be explored, including more elaborate priors such as generalised total variation, Plug-and-Play (PnP) methods, and end-to-end deep neural networks (DNNs). Further extensions may include more general forward models that incorporate spatially varying blur and ghost-image intensity, spatially localised ghost-image formation (e.g., confined to foreground or background regions), and other physically consistent image formation effects. 

\section*{Disclosures}
The authors declare no conflicts of interest.


\bibliography{refs}

\appendix
\section{Derivation of the source-limited ghost-image blur}
\label{app:blur_spot}
\begin{figure}[htbp]
    \centering
    \includegraphics[width=0.8\linewidth,trim={7 20 7 20}]{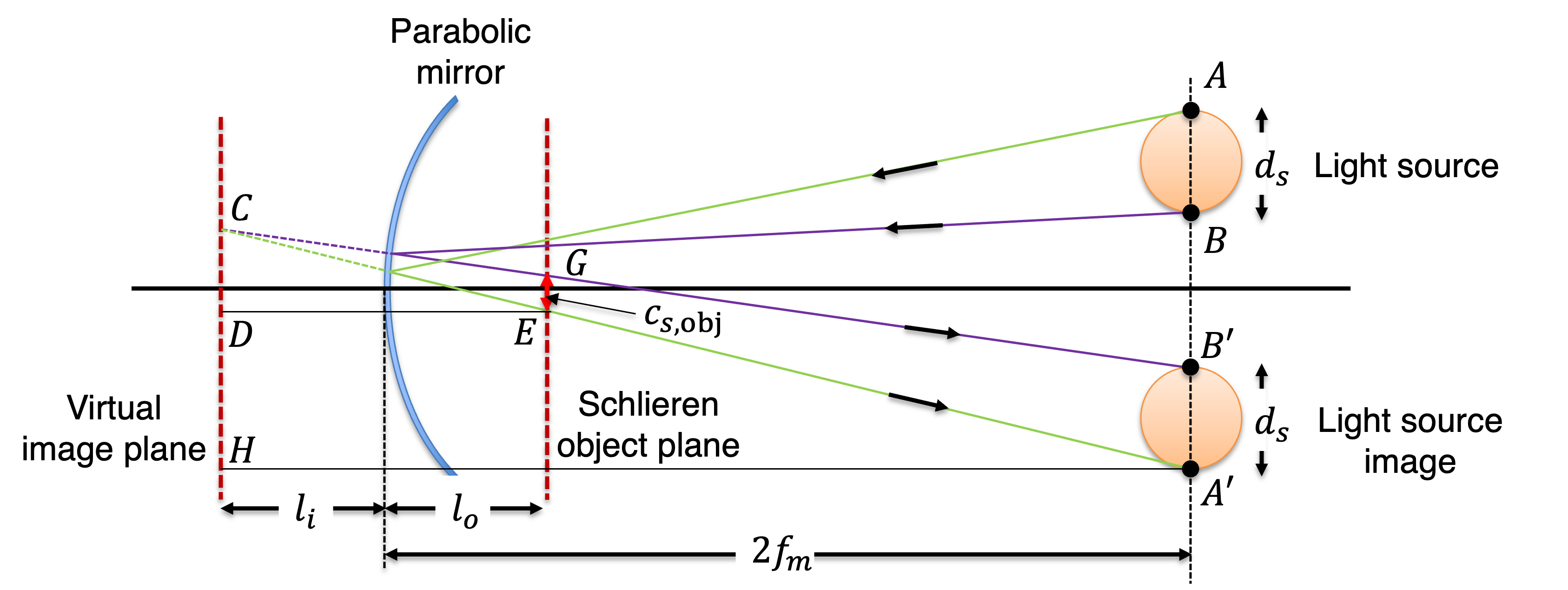}
    \caption{Ray geometry for the derivation of the source-limited ghost-image blur.
    }
    \label{fig:source_blur_derivation}
\end{figure}

The source-limited blur of the ghost image arises from the finite aperture of the light source. The ray geometry used to derive this contribution is shown in Fig.~\ref{fig:source_blur_derivation}.

The light source aperture is located at a distance \(2f_m\) from the parabolic mirror. Therefore, 
the mirror forms an image of the source at the same axial distance \(2f_m\), with 
magnification \(m_s=-1\). Hence, the source aperture \(AB\) and its inverted image \(A'B'\) have the same diameter,
$
|AB|=|A'B'|=d_s.
$ 
Rays originating from \(A\) and \(B\) are reflected by the mirror to form the corresponding image points \(A'\) and \(B'\), respectively. To determine the blur produced by a point \(C\) of the virtual ghost image, we consider the limiting reflected rays whose backward extensions intersect at \(C\). These limiting rays correspond to the extreme points of the finite source aperture and therefore bound the transverse extent of the light at the Schlieren object plane. We denote the diameter of this blur 
spot by $c_{s,\mathrm{obj}}$. From the similar triangles $CEG$ and 
$CA'B'$, we obtain

\begin{equation}
    \frac{c_{s,\mathrm{obj}}}{d_s}
    =\frac{|GE|}{|A'B'|}=\frac{|CE|}{|CA'|}.
    \label{eq:supp_similar_triangles}
\end{equation}

From the similar triangles $CDE$ and $CHA'$,

\begin{equation}
    \frac{|CE|}{|CA'|}
    =
    \frac{|DE|}{|HA'|}
    =
    \frac{l_o+|l_i|}{2f_m+|l_i|}.
    \label{eq:supp_similar_triangles1}
\end{equation}

Combining~\eqref{eq:supp_similar_triangles} and
\eqref{eq:supp_similar_triangles1} gives
\(c_{s,\mathrm{obj}}=d_s(l_o+|l_i|)/(2f_m+|l_i|)\).
The virtual-image distance is related to the Schlieren-object distance
through the mirror equation,
$1/f_m = 1/l_o - 1/|l_i|$, which gives
$|l_i| = f_m l_o/(f_m-l_o)$.
Thus,

\begin{equation}
    c_{s,\mathrm{obj}}
    =
    d_s\frac{
        l_o+\dfrac{f_m l_o}{f_m-l_o}
    }{
        2f_m+\dfrac{f_m l_o}{f_m-l_o}
    }
    =
    \frac{d_sl_o}{f_m}.
    \label{eq:supp_substitution}
\end{equation}

Since the camera is focused on the Schlieren object plane, the blur spot is imaged onto the sensor with primary-image camera magnification \(m_p\), giving the corresponding blur-spot diameter at the sensor \(c_2=m_pc_{s,\mathrm{obj}}=d_sl_om_p/f_m\). Converting this diameter to a radius in pixel units using the effective pixel pitch \(p_{\mathrm{eff}}\) gives
\[
r_2=\frac{c_2}{2p_{\mathrm{eff}}}
=\frac{d_sl_om_p}{2f_mp_{\mathrm{eff}}},
\]
which is the expression used in~\eqref{eq:source_limited_pixel_blur} of the main manuscript.






\end{document}